\documentstyle[aps,multicol,eqsecnum,epsf]{revtex}

\def\H{{\cal H}}
\def\E{\epsilon}
\def\ept{{\tilde \epsilon_1}}
\def\tz{{\cal Z}}
\def\x{{\bf x}}

\def\z{{\bf{\hat z}}}
\def\r{{\bf r}}
\def\be{\begin{equation}}
\def\ee{\end{equation}}
\def\df{D}
\def\nb{{\bf \nabla}}
\def\u{\phi}
\def\lepsilon{\epsilon}

\def\h{{\bf h}}

\def\v{{\tilde {\bf h}}}

\def\t{{\tau}}

\begin{document} 
\title{Finite Size Effects in Vortex Localization} 
\author{Nadav M.  Shnerb}
\address{ Lyman Laboratory  of Physics, Harvard  University,
Cambridge, MA 02138}
\date{\today}
\maketitle

\begin{abstract}
The equilibrium properties of  flux lines pinned by columnar disorder 
are studied,  using the  
analogy with the time evolution 
of a  diffusing scalar density in 
a randomly amplifying medium. Near $H_{c1}$, the 
physical features of the vortices in the localized phase 
are shown to be determined by  the  density of states near the band 
edge. As a result,  $H_{c1}$ is inversely proportional to
the logarithm of the sample size, and the screening length 
of the perpendicular magnetic  field  {\it decreases} with 
temperature. For large tilt the extended ground state 
turns out to wander in the plane perpendicular to the 
defects  with  exponents corresponding  to a 
directed polymer in a  random medium,
and the energy difference between two competing metastable states 
in this case is extensive. The divergence of the effective potential 
associated with strong  pinning 
centers as the tilt approaches its critical value 
is discussed as well. 
\end{abstract}

\begin{multicols}{2}
\section{Introduction}

The physical   properties of  vortex lines 
in the mixed phase of type II superconductors has  
become subject of  intense research in recent years \cite{rmp}. 
When an external current density is applied to the bulk of the system, 
the flux lines may  start to move under the action of  the 
Lorentz force. Within 
a perfectly homogeneous system this driving Lorentz force is balanced only
 by the 
friction force opposing the 
steady state velocity of the flux lines, so that there 
is a dissipation coupled to  
the appearance of a  finite electric field  which results from  
the flux motion. To avoid this effect, flux lines in the mixed phase 
should be  pinned by  inhomogeneities in the underlying crystal 
structure, such as the point  
defects associated with vacancies of oxygen atoms in 
high temperature cuprates \cite{point}. It turns out that the 
pinning of the
vortices is much stronger when these impurities are in the form of 
correlated disorder, like  twin boundaries 
or columnar defects\cite{civle,twin},  
aligned along the direction of the external magnetic field.
If the thermal fluctuations are small enough, 
the flux line may lie along the entire  extended defect in
the bulk; 
this is in contrast  to  short scale disorder pinning, in which the 
line should accommodate itself to the potential fluctuations and  hence 
increases its elastic  energy. However, 
the correlated defects pinning becomes less effective when  the 
direction of the external magnetic field is tilted with respect to the 
anisotropy axis, which we take to be along the ${\bf{\hat z}}$ direction.  
At some critical tilt, for which the free energy per unit length of the defect 
is less than the free  energy associated with the perpendicular field, a 
pinning-depinning   phase transition occurs and 
the flux lines are  delocalized. 

The static and dynamic response of the  flux lines 
in the presence of columnar defects  have been 
considered by Nelson and Vinokur \cite {nel-vin}. Using the mapping of 
flux lines in $d+1$ dimensional superconductor to the world lines of bosons
in a $d-$dimensional quantum system, the authors 
 identified  the phase space diagram of 
the system which contains  
a high temperature ``superfluid'' and low-temperature
``bose glass'' phases, as well as Mott insulator at the matching field, 
$B_\phi = n_{pin} \phi_0$, for which there is one flux line per defect. 
At low temperature, this matching  field separates the ``dilute'' 
region of the bose glass
phase, for which the vortex lines are pinned individually by the defects,
from the high density region, where interactions are important in 
determining the localization length and transport properties of the flux 
lines.

In the low field region, the  vortices are localized by the interaction with
the correlated defects. Each pin  is the analog of a $2D$ potential well
which may be described  (up to logarithmic corrections) as a 
cylindrical square well. The temperature of the superconductor, in
turn, corresponds to the Plank constant $\hbar$ of the quantum boson system. 
For the dilute vortex arrays, where the pinning energy is larger than the 
interaction energy, there are two regimes. For low temperature 
the localization length 
is given approximately by  the radius of the defect so that each flux line 
is localized by $\it{ one}$ defect. As the temperature increases,  the 
localization length of one defect grows exponentially with $T^2$, and the 
flux line is then localized by several defects, forming an effective $d$ 
dimensional potential well in the corresponding boson system. 

The depinning  of  flux line which occurs  as a result of external field  tilt 
has been carefully investigated by Hatano and Nelson \cite {hat}. The   
Hamiltonian of the corresponding boson problem is no longer Hermitian;
the kinetic term of the Hamiltonian  is subject to an 
imaginary gauge transformation  where the gauge field  $\h$ is related to the 
perpendicular magnetic field ${\bf H}_\perp$ via $\h = {\bf H}_\perp \phi_0
/(4 \pi)$, where $\phi_0$ is the flux quantum. 
In the absence of tilt, the 
probability of finding a point of the vortex at transverse displacement 
$\r$ relative to the center of the pin is independent of $\z$, and given
by the square of $\phi_{gs}(\r)$, the ground-state wavefunction
of the Hermitian Hamiltonian. For small tilt,     
there are left and right  ground-state eigenfunctions which 
correspond to the  left and right
``tilting'' of the  Hermitian ground-state are still localized.
It turns out that the flux line is described by these 
left and right eigenfunction at the bottom and the top of the 
sample, while deep in the bulk the probability function 
approaches its  Hermitian limit.     
Typically, the ``surface roughness'' associated with the tilt extends
 into the bulk up to some  characteristic distance 
which diverges as the 
tilting angle approaches the critical angle, for which the flux line 
delocalizes and the current response becomes linear.
 
This non-Hermitian delocalization process has been shown to be
of importance for  other physical systems as well. In general, 
the time evolution of diffusing  scalar field  in  random environment 
is determined by the eigenstates of Liouville operator which is the 
analog of the  Hermitian Hamiltonian of a quantum particle in a disordered
system; the effect of convection may be modeled by the 
imaginary gauge transformation. The appearance of  
extended states as convection is increased beyond a critical value,
associated with the  complex spectral points  of the 
resulting  non-Hermitian  operator,  is the manifestation of the 
delocalization transition. The dynamics of  strongly driven charge 
density waves \cite{cdw}, and the growth modes of biological 
populations \cite{us}, are among the systems considered.

In this paper we study this tilt induced delocalization 
transition at very dilute flux 
line concentrations, i.e., near $H_{c1}$. We assume, therefore, 
that only the low-lying free energy states of the system are occupied, 
and that the repulsive interaction between vortices may 
be taken into account by sequential filling of energy levels. 
Although this assumption is self consistent in the localized phase, 
it turns out that it is valid in the extanded phase only 
if the sample is finite. If the  size of the  sample is taken 
to infinity at finite vortex density, interaction may 
lead to a smecticlike flux crystal, as has been discussed in Ref. 
\cite{nel-vin}.

This paper is organized as follows. In Section 2, a brief introduction 
to the boson - vortex analogy is presented, as well as the correspondence 
between the statistical mechanics of  flux lines and the time evolution of 
a  scalar field. Section 3 deals with the finite size effects  in the 
localized phase;  both the critical field  $H_{c1}$ and the penetration depth 
of the perpendicular field are shown to be determined by the statistics of 
the tail of the density of states. Since this function is strongly related to
 the size of the system,
physical properties of the pinned vortex 
are determined by the actual size of the bulk. 
In Section 4 we consider the delocalized phase,  in which 
we recognize that the low lying states are extended and are related to the 
properties of  Burgers's equation with 
conservative noise in $1+1$ dimension \cite{FNS}. Some technical 
details are given in the appendices.

\section {Flux line in columnar disorder}

We review  here  the basic physics of a single 
flux line in $d+1$ dimensional superconductor, as has been 
discussed in \cite{nel-vin} and \cite{hat}.

\begin{minipage}[t]{3.2in}
\epsfxsize=3.2in
\epsfbox{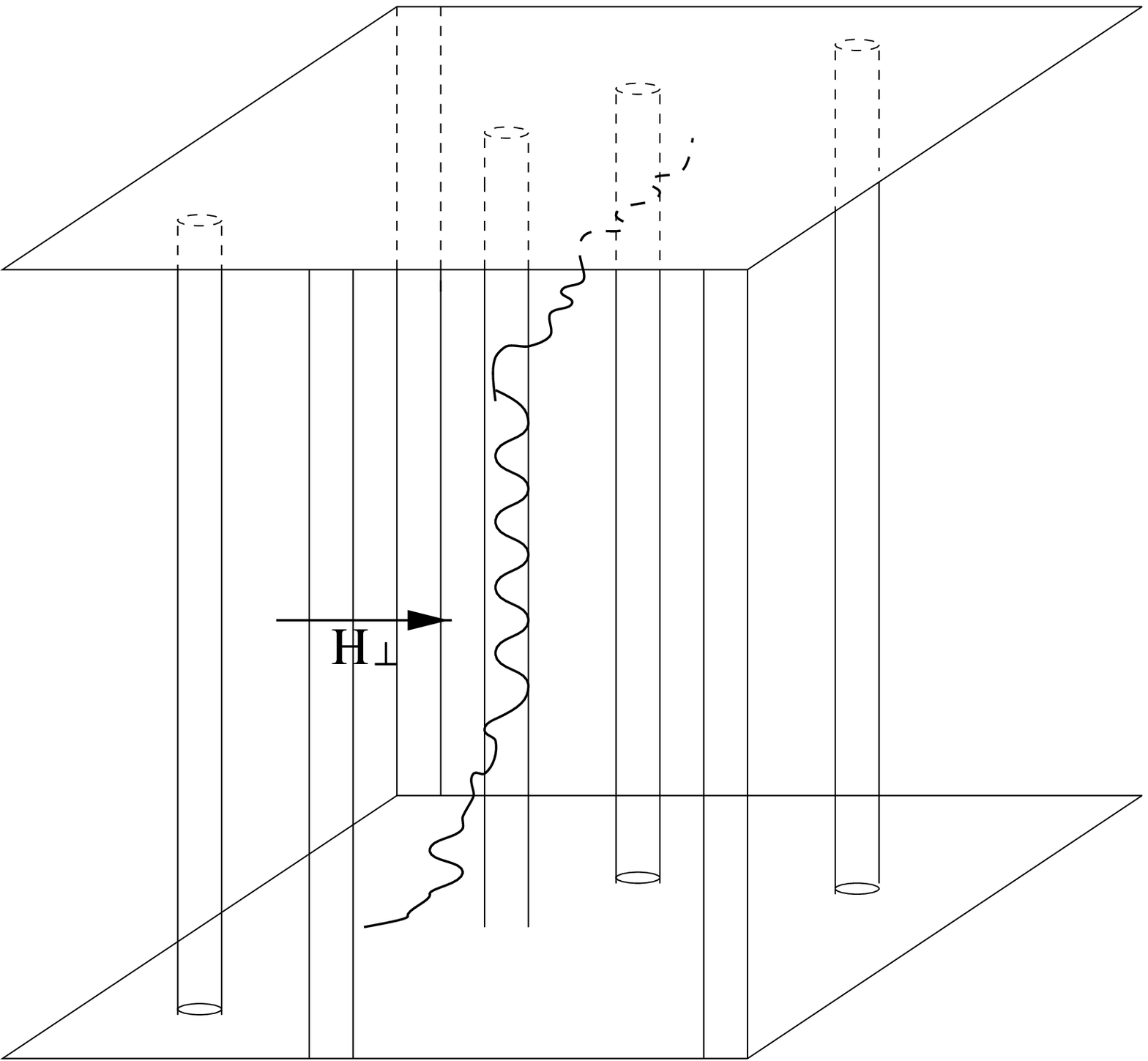}
\begin{small}
FIG.\ 1. 
Vortex line in a superconductor with columnar disorder.
If $H_\perp=0$, the flux line is localized, i.e., trapped 
by one or few pins into some region in the plane 
perpendicular to the correlated disorder. As the external 
magnetic field is tilted away from the columns, the 
flux line tends to delocalize and tilt in  the 
direction of the external field.
\end{small}
\vspace{0.2in}
\end{minipage}

Consider a flux line in a superconducting  sample of thickness $L_0$
when the   sample is pierced by columnar pins, such as long alighted 
columns of
damaged material, as illustrated in  Fig. 1.

Neglecting overhangs of the line, we define the flux line by its trajectory 
${\r(z)}$. 
The free energy of this line may be written as, 
\begin{eqnarray}
F &=& {\ept \over 2} \int_0^{L_0} ({d\r(z) \over dz})^2 + {1 \over 2}
 \int_0^{L_0} [a + U(\r(z))] dz \\ 
 & \  -& {{\bf H_\perp} \phi_0 
\over 4 \pi}  \int_0^{L_0}
({d\r \over dz})dz \nonumber
\end{eqnarray}
where $\ept$ is the tilt modulus of the flux line, i.e., the energy of 
the flux line per unit length, and the elastic contribution 
$ {\ept \over 2} ({d\r(z) \over dz})^2$ is the first nontrivial 
term in the  small tipping angle expansion of the line energy of a  nearly
straight vortex line. $U(\r)$ is the random potential which  arises  
from a $z$-independent set of disorder-induced columnar pinning potentials,
and ${\bf H_\perp}$ is the perpendicular field.  $U(\r)$ is taken from 
some bounded distribution, such as square distribution of 
width $\Delta$ around zero; its bias, $a$, is determined by 
$\ept$ and the parallel  component of the external magnetic field, i.e., 
$a = ( \ept - {H_\|\phi_0  \over 4 \pi})$. Note that we assume 
$H_\perp$ to be much  smaller than $H_\|$, 
hence we neglect terms of order $H_\perp^2$ in (2.1). 

Using the analogy between the partition function for this flux line, 
$Z = e^{-F/T}$, and the path integral formulation of quantum mechanics, 
it is easy to see that the conditional probability 
$\tz (\r,z ; \r_0,z_0)$
obeys  the Schroedinger-like equation,  
\be\label{res1}
-T {\partial \tz  \over \partial z} = 
- {T^2 \over 2 \ept} \nabla^2 \tz  + [a +  U(\r)] \tz  - 
{T \over \ept} \h_\perp \cdot \nb \tz
\ee
where $\h_\perp \equiv  {{\bf H_\perp} \phi_0 \over 4 \pi}$ is the 
dimensionless perpendicular field. 

Let us now  map this  flux line model, which satisfies
Eq. (\ref{res1}), onto  a model for the diffusion of a  
passive scalar density, 
$\phi (x)$, suppressed/amplified  by a quenched random potential, and driven 
by an  external drift. The  time evolution of such a  system 
is described  by the  equation:

\be\label{1}
\partial_t \phi(\r,t)  = D \nabla^2 \phi(\r,t) - [b + V(\r)]  \phi(\r,t)
+  \v \cdot \nb \phi(\r,t), 
\ee
where $V(\r)$ is
 the   random amplification/suppression rate, and $D$ is the diffusivity. 
Clearly, the partition function of the flux line  satisfies Eq. (\ref{1}) for
$\phi$, where $D$ plays the role of temperature divided by the tilt modulus,
$D = T/(2{\tilde\epsilon}_1)$, $\v$ is proportional to the 
perpendicular magnetic  field, $\v ={\h_\perp \over \ept} $ and 
the columnar disorder
in the superconductor,as well as its bias,  are
normalized by the temperature, i.e., $V(\r) =  U(\r)/T$  and 
$b = a/t$.  The correspondence between the various scalar field and 
magnetic vortex quantities are summarized in table 1. 
\begin{center}
{\bf \qquad \qquad Table 1}\newline 
\vskip 0.5mm
\begin{tabular}{|c|c|}
\hline Scalar Density   & Flux Line  \\\hline \hline
\hline $D$ & $T/2\ept$ \\\hline
$V(\r)$ & $U(\r)/T$ \\\hline
$\v $ & $\h_\perp/T $\\\hline
$b$ & $ (\ept - {H_\| \phi_0  \over 4 \pi}){1 \over T}$ \\\hline
$\phi$ & $ \tz$ \\\hline
\end{tabular}
\end{center}
\vskip 2mm
Eq. (\ref {1}) may be written as $\partial_t \u = - \H \u$, 
where $\H$, the Hamiltonian,  is the  linear operator  which 
generates the time  evolution of $\u$,
\be\label{ham}
\H = - D \nabla^2  +  [b + V(\r)] -  \v \cdot \nb.
\ee
The dynamics of this system is
determined by the eigenvalues and the eigenvectors of the Hamiltonian.  
For $\v = 0$, the  operator,  
$\H = -  D \nabla^2 + [b+ V(\r)]$, 
is Hermitian and it is well known that, for strong 
enough disorder, all its eigenfunctions are real and localized; the 
localization length is maximal at the center of the energy band, and minimal
at the tails \cite{el-st}. 
When $\v \neq 0$, the above Hamiltonian  is no longer Hermitian, and 
it may be diagonlized using a system of  left and right eigenvectors. 
Since the  perpendicular field   
term may be absorbed into the Laplacian by completing the square, $\nb 
\to \nb + {\v  \over 2D}$,  the  right and left eigenfunctions
of the new Hamiltonian  
are related to the eigenfunctions of $\H(\v = 0)$  via an  imaginary 
gauge transformation;
if $\u_n^{\v = 0}(\r)$ is an eigenfunction of the Hermitian problem, then 
\begin{eqnarray}\label{gauge}
\u^{R}_{n;\v}& = & e^{\v \cdot \r / D} \u_{n,\v = 0}   \\  \nonumber
\u^{L}_{n;\v}& = & e^{-\v \cdot \r / D} \u_{n,\v = 0}
\end{eqnarray}
are the  eigenfunctions of the non-Hermitian operator with the {\it same}
eigenvalue, 
provided that $\xi_n$,  the  localization 
length   in the non-driven problem, is less than $D/{\tilde {h}}$. Thus, for small 
tilt, there is a {\it spectral rigidity} - the eigenvalue spectrum 
is robust. When  $\v$ increases, the eigenfunctions  become extended and 
the boundary conditions of the sample should be taken into account. Since 
(\ref {gauge}), in general, does not satisfy these boundary conditions, the 
new eigenfunctions are no longer   related  to the $\v = 0$ case by a simple
gauge transformation. As a result, the eigenvalues are changed; in case of 
periodic boundary conditions, complex eigenvalues appear when 
$D/{\tilde {h}}$ becomes 
smaller then $\xi_n$. As $\v$ is  increased, these delocalized states appear
first at the band center, for which the localization length is maximal,  
then move outwards, as has been discussed in \cite{hat}.

The mapping of  Eq. (\ref{res1}) to
Eq. (\ref{1}) implies that for thick samples, $L_0 \to \infty$, 
the thermally excited flux line relaxes into the spatial configuration
equivalent to the ground state eigenstate of  
$\H$. For small $\h_{\perp}$, this state is localized, so that the
flux line is localized around some spatial point at the bulk of the sample. 
This phenomenon is known as the  transverse Meissner effect. As has 
been shown by \cite{hat}, the ends of the pinned flux line begin to 
tear away from the pinning center as the tilt increases. The ``penetration 
depth'' of the perpendicular field, associated with the width of the 
region near the surface in which the transverse Meissner effect breaks down, 
diverges as the tilt approaches $\v_c$. Above this value, 
the eigenstate delocalizes, and
the resulting trajectory  wanders  across 
the  entire sample. 

\section{Magnetic response in the pinned phase}
\subsection{Size dependence of  $H_{c1}$}
The lower critical field, $H_{c1}$, is defined in a random system 
 as the minimal field 
in which the first vortex line enters the sample. One could easily 
recognize from Eqs. (2.1-2.4) that the condition for 
one flux line in  the sample to be energetically favored over the
completely diamagnetic phase, is that the Hamiltonian (\ref{ham})
admits negative eigenvalues; the spectral points   of (\ref{ham})
are  proportional to  the energies per unit length of the corresponding 
configurations of the vortex.  $H_{c1}$ depends, thus,  on the ground 
states of the Hamiltonian;  for random potential  
one should consider the 
statistical properties of  the lower  tail of the density of  states. 
Note that we consider here the physical  
case of {\it bounded} disorder, 
such that for each sample there is a well defined 
lower critical field. This is in contrast to the discussion 
given by Larkin and Vinokur, who address the question of magnetic response 
for  unbounded, Gaussian random potential in an  
infinite sample. In such a case there is 
no  well defined $H_{c1}$ \cite{larkin}.

For a  statistically homogeneous sample, any rare fluctuation 
is realized as one takes
the bulk  size to   infinity. In particular, one may finds any large region 
in which the potential is close to its lower bound, so for 
infinite sample the minimal value of the potential 
$V(x)$ sets the magnitude of $H_{c1}$, i.e., 
\be 
H_{c1}^{\infty} = {4 \pi (\ept -  
\Delta) \over \phi_0}\; .
\ee  
For finite sample this is not the case. Typically,
as the sample size increases, $H_{c1} \to H_{c1}^{\infty}$; however, 
the form of the size dependence  of the critical magnetic field is not trivial.

Consider, for example, a hole in the superconductor, i.e., 
a region of linear size $L$ where the elastic energy 
of the flux line vanishes, $\ept = 0$. The free energy of a vortex line 
which is trapped in this hole comes entirely from the limits on its 
transverse motion sets by the boundaries of this hole. Far away 
from the bulk critical temperature (i.e., when  $\ept$ in the bulk  is much 
bigger than the typical free  energy per unit length associated with the 
trapped flux line) one may assume no vortex displacement 
along the hole boundaries, and the situation corresponds to 
a quantum mechanical particle in a box. In such a case the 
tail of the spectrum of (2.4) looks like
\be
\epsilon(k) \sim  D k^2,
\ee   
where  the minimal value of $k$ is  sets by the system linear  size $L$, 
$k_{min} \sim 1/L$, so that,    
\be
H_{c1} \sim     {4 \pi T^2  \over \ept  \phi_0 L^2}.
\ee
This result reflects the fact that the entropy of the vortex line decreases
if it is  confined  to finite region, hence its free energy per unit length 
increases.  At zero temperature this finite size correction to $H_{c1}$
disappears; as the temperature approaches its critical value, 
the potential steps at the hole boundaries become small, and in 2D 
this contribution to the free energy vanishes exponentially.     

Let us consider now the disordered case.  In Appendix  A we show that for a
simple  lattice model with square distribution of the disorder
the density of states per unit volume  near the minimal possible value of the 
energy (which we set, for convenience, to zero since this  value
 is absorbed  in the definition of  $H_{c1}^{\infty}$) is
\be
g(\epsilon)   \sim  e^{-\left({D \over l_0^2  \epsilon} \right)^{d/2}}
\ee  
where $  l_0 $ is the lattice constant. One may estimate the 
energy difference $\Delta \epsilon$ between the ground state and 
zero for a typical sample of linear size $L$  by the condition
\be
\left({L \over l_0}\right)^d g(\epsilon)\Delta \epsilon \sim 1
\ee
so for large samples the ground state energy  is   given by
\be
\epsilon_0(L) \sim  {D \over l_0^2  \ln^{2/d}(L/l_0)}.
\ee
and 
\be
H_{c1} \sim  H_{c1}^{\infty} +{4 \pi T^2  \over  \ept \phi_0 l_0^2 
 \ln^{2/d}(L/l_0)}.
\ee
In contrast to the hole example, one sees that the  critical parallel
field approaches 
$H_{c1}^{\infty}$ much slower, i.e.,  finite size effects are much stronger
in the presence of disorder. 
For the   
tilt field below its critical value the situation remains the same, since 
the spectrum does not change. 

Upon taking reasonable parameters for high temperature superconductors, such as
 $T \sim 70^0 K$, $l_0 \sim 100-1000 \stackrel{\; \; o}{A}$ and 
$\lambda \sim 1000 \stackrel{\; \; o}{A}   $, one 
finds, however, that the entropic corrections to $H_{c1}^{\infty}$ are
very small, $ {\delta H_{c1}/H_{c1}}    \sim   10^{-5}$. Thus, for 
almost any realistic  sample size  
one could take the value of $H_{c1}^{\infty}$ as the 
lower critical field. Since in most cases the radius of the columnar pins
is comparable to, or  bigger than, the superconducting coherence length,
the range of the  external field corresponding 
to the  Meissner phase turns out 
to be very  narrow. For other physical situations when the time evolution 
of  some scalar density  is given by (2.3), these finite  size corrections may 
by of importance.

\subsection{ Transverse Meissner Effect for Dilute Concentration}
Given a system described by the equation
\be\label{liouvill} 
\partial_t \u  = - \H \u
\ee
where  $\H$ is, in general, a time independent non-Hermitian  
operator with a complete set of left and right 
 eigenstates $\u_{n,L},\u_{n,R}$ with the corresponding  
eigenvalues $\{ \epsilon_n \}$ (i.e., the complex  
``energy spectrum''
of the possibly non-Hermitian operator),  the time evolution of  
the  normalized amplitude with initial condition  $\u(\r,0)$
is given  by its  spectral decomposition:
\begin{eqnarray}\label{prob}
{\hat \u} &\equiv& {\u(\r,t) \over \int d\r \;  \u(\r,t)} 
\\ 
& \; =&{ \sum_n <\u_{n,L}|\u(\r,0)> \u_{n_R}(x) \exp(-\epsilon_n t) 
 \over \int d\r \sum_n <\u_{n,L}|\u(\r,0)> \u_{n_R}(x) \exp(-\epsilon_n t) }
\nonumber
\end{eqnarray}
where $<..>$ is defined as the inner product 
$<\psi|\phi> \equiv \int \  d\r
\  \psi^*(\r) \  \phi(x)$. 
At long times, the system is  dominated by the ground state, i.e., 
the state for which the real part of $\epsilon_n$,  $ Re \; \epsilon_n$,
 is {\it minimal}. Using the mapping between the partition function 
of the vortex system and the scalar field $\u$, one observes that Eq. 
(\ref{prob}) describes the surface  roughness of the flux line: although deep
in the bulk, which is  the equivalent to the 
``long time'' of (\ref{prob}), 
the line is stuck to the  spatial configuration corresponding to the 
ground state, this is not the case near the surface, where the  ``short time''
limit of (\ref{prob}) is relevant \cite{hat}.

Let us discuss  this phenomenon for the  $\v = 0 $ case, 
when  all
the wavefunctions are localized. Spectral decomposition of  small, 
localized, initial conditions 
near the most rapidly growing state requires projection 
upon the eigenvectors 
of the Hamiltonian. These projections, in general, 
decrease exponentially with the 
distance between the fluctuation center and the localized eigenvector.
For example, if the initial 
 fluctuation is  centered around zero, the projection on an eigenstate 
localized at $\x$ is
\be 
<\u_{\x=0}(\r,0)|\u_{\x}(\r)> \sim \exp^{- \kappa_n |\x|},
\ee
where 
$\kappa_n$ is the inverse  localization length of the n-th eigenstate.   
The fastest growing states of $\H$ dominate the system at long times; 
at any finite
time, however, there is a competition between the initial conditions and the
growth rates, which we explore using the following model.

Assume for simplicity that  
the localization length is constant for all states of the 
system, i.e., it is independent of the eigenenergy as well as the 
spatial position.
The probability amplitude   $\u_{\x}$  
generated by  a localized  initial condition 
$\u_{\x=0}(\r,0)$  
is proportional to
\be\label{sad}
{\u_{\x}(t) \over \u_{\x=0}(t)} 
\sim e^{- \kappa |\x|} e^{[\lepsilon_{\x}- \lepsilon_{\x =0}] t},
\ee
where  $\lepsilon_{\x}$ is the eigenvalue of a growth mode  localized
at $\x$. 
We shall study the relaxation of the first moment of $\u(\x,t)$,
\be \label{moment}
{\bar \x}(t) = \int d\x \; \x \; {\hat \u}(\x,t) = 
{\int d\x \; \x \; \u(\x,t) \over 
\int d\x \; \u(\x,t)}
\ee
to the value $\x_{gs}$ it assumes at long times,
\be \label{mom2}
\x_{gs} = \int d\x \; \x \;  \u_{gs}(\x)
\ee
where $\u_{gs}(\x)$ is the lowest energy eigenmode 
of the Hamiltonian. Given  
$\lepsilon_{\x}$, one can  optimize (\ref{sad}) in order 
to find the location of the  
eigenstate which dominates the system at finite time. For 
a finite sample it is clear that as $t \to \infty$ the 
ground state (lower free energy) mode localized at position
$\x_{gs}$ will dominate. The long time  
convergence of the normalized amplitude
to the ground state location is also given by (\ref{sad}), where 
we use a  normalization with respect  to the 
final location  amplitude,  i.e., 
\be\label{sad1}
{\u_{\x}(t) \over \u_{\x =\x_{gs}(t)}} \sim 
e^{- \kappa |\x_{gs}|} e^{- (\lepsilon_{\x = \x_{gs}} 
-\lepsilon_{\x}) t}.
\ee 

In random samples one may assume that the eigenenergies 
are described by a statistically homogeneous density of states
$g(\epsilon)$, and a typical spatial distance 
$R = |\x - \x_{gs}|$ associated with energy difference 
$\Delta \epsilon = \epsilon_{\x_{gs}} - \epsilon_{\x}$
is given by the analog of Eq. (3.6) \cite{el-st}

\be\label{unif}
\Delta \lepsilon \  R^d \  g(\epsilon) = 1.
\ee

A similar mapping into variable range hopping in semiconductors
has already been exploited in \cite{hat} for
finite density of vortices  in the superconducting medium.  
The long range repulsive   interaction between flux lines has been taken 
into account, in 
the strongly  localized regime,  by forbidding 
multiple occupancy of localized states, and   the ``ground state''
is defined by filling the energy band up to the ``Fermi surface'' determined
by the chemical potential of the particles.  If all  the states are 
localized and the filling factor is such that the 
Fermi surface is not at the tails of the band, the density of states  
at the chemical potential, $g(\mu)$, may be taken as a constant 
independent of the 
energy. Upon 
substituting $\Delta \lepsilon(R) =1/ (g(\mu) R^d)$ into  Eq. (\ref{sad1}), 
one can  study the long time convergence to the ground state location, in terms
of the 
distance $R^*$ away from $\x_{gs}$ for which the normalized amplitude 
is maximal. One then finds  that the approach of $|\x - \x_{gs}|$
to zero at long times is given by a stretched exponential \cite{hat}, 
\be \label{decay1}
|\x - \x_{gs}|   \sim \exp[-({t \over t^*})^{1 \over d+1}],
\ee
where $\t^*$ is the  characteristic time of the 
decay, $t^* =  g(\mu)/\kappa^d$. Since the time in Eq. (2.3) corresponds
to the the length in the correlated direction of the flux problem 
normalized by the temperature,   the  penetration depth 
of the perpendicular magnetic field is $T  g(\mu)/\kappa^d$.

For a very dilute concentration of flux lines, 
the ground state (to which the system eventually evolves), 
is in the lower  tail of the 
density of states and is {\it always}
localized, even for weak disorder in high 
dimensions. The nature
of the surface roughness 
is determined by the other low energy states at the tail of the 
density of states  function.   Since the ground state 
is now near  the bottom of the band, the assumption that the density of
states in this region 
is energy independent is no longer valid. One may estimate the 
time needed for the decay of a state in the tail to the 
ground state by optimizing $(\ref{sad1})$ with respect 
to $R$, using (\ref{unif}) and  (3.4). 
Taking the logarithm of both sides of (\ref{unif}), 
we get,  up to logarithmic corrections, 

\be
\Delta \E \sim {\df \over (\ln R )^{2/d}}.
\ee

From this we obtain  the semiclassical expression for 
the distance away from the ground state at time  $t$:

\be\label{tail}
|\x - \x_{gs}|  \sim \exp [-{{t/t^*} \over 
\log{(t/t^*)}^{1+2/d}} ]         
\ee
where now $t^* = l_0^2 /D$, or in the vortex language, the width
of the  surface roughness 
is  proportional to  $\ept l_0^2 / T$. Note that the effect of temperature
(or the diffusion constant) 
in this expression is to {\it decrease} the surface roughness rather then 
increasing it as in the former case; the reason for this 
counter-intuitive result is that the  temperature dictates 
the energy spacing at the tail. Lower temperature implies bigger 
spacing, so that the effect of neighboring sites on a pinned flux line
becomes smaller. Of course, the width of the tail itself is proportional 
to the temperature, so that the result (\ref{tail}) becomes 
meaningless at the limit $T \to 0$.

This result implies that at small concentrations, i.e., close to 
$H_{c1}$, the localization of flux lines is more effective than 
in the case considered by \cite{hat}. Samples which are not  thick enough 
to localize the vortices at high magnetic field may   do so near 
the lower critical field, when the effects of tail statistics are of 
importance.

For a finite but small tilt  
$\v$, the low-lying states are still  localized, and  the 
situation is essentially the same. At short times, there can be transition 
effects due to  
extended  states which  overlap with the initial condition, but
at long times the normalized amplitude is localized into the 
ground state. For strongly localized states near the band edge
the tilt  can  be absorbed into  
the gauge transformation (\ref{gauge}), the only change introduced 
in the consideration above is that the inverse localization length 
$\kappa$ becomes anisotropic, 
\be\label{loc}
\kappa (\v) = 
\kappa(\v = 0)   - \v.
\ee 
As a result, one should 
optimize (\ref{sad}, \ref{sad1}) with respect to $\theta$, the angle between
$\v$ and $\x$. The result  turns out to be  equivalent to   
replacing  $\v$ in (\ref{loc}) by its absolute value $\tilde{h}$.

\section{extended states at large tilt }
\subsection{Delocalized flux line as directed polymer}

Let us consider now the limit of very large tilt, in the sense that 
even the ground
state of our  system is  delocalized, i.e., ${\tilde {h}} >> D/\xi_0$.  
As $\tilde{h} 
\to \infty$, the external magnetic field is perpendicular to the columnar 
defects; the flux line is now free in the $\z$ direction, while in the 
$xy$ plane it sees the cross section of these defects, i.e., point disorder. 
One might then guess   
that the flux line freely  diffuses in the $z$ direction,
while in the $xy$ plane it looks like a directed polymer in  a $1+1$
dimensional  random medium, 
that is, the projection of the flux line on the $xy$ plane is a pinned 
string characterized by the wandering exponent $1/z = 2/3$ \cite{FNS}. 
It is separated 
from  competing metastable states by energy barriers which 
(up to logarithmic corrections) scale like $L^\beta$, where $\beta = 1/3$ 
and $L$ is the linear system size  \cite{FNS,huse}.

The same features  characterize the most rapidly growing
eigenmode  of the Hamiltonian (\ref{ham}). 
The  low lying eigenstates of the non-Hermitian  
Hamiltonian  are extended
in the tilt direction while wandering in the transverse direction 
with the same
exponents $1/z$ and $\beta$. Suppose $L$ is the length
of the sample in the $\v$ direction, and  $W$ is the width in the 
perpendicular direction, measured in units of some short length cutoff. 
For $W >> L^{2/3}$ there are many, spatially 
uncorrelated,   competing 
metastable states in the sample;  the decay of such a  
state into the ground state involves  activation
above a potential barrier of order $L^{1/3}$ \cite{kardar}, 
so that the time needed 
for such a process is of order $t_0 \sim \exp(L^{1/3})$. On the other 
hand, as $W<L^{2/3}$, there is only one  effective low energy state; 
since the energy  barriers of the  directed polymer are equivalent to the 
surface roughness in KPZ model, the potential barriers 
scale as $W^{1/2}$ in this case. In general, one sees that, in $2+1$ 
dimensions,  when
the low-lying states are extended, the time needed for the relaxation to the 
ground state is macroscopic and proportional to the (longitudinal or 
transverse) size of the system.

In order to quantify these considerations, we may map Eq. (2.3) into a 
nonlinear problem with additive disorder using  
Cole-Hopf transformation, i.e.,
\be 
\u(\x,t) = \exp^{( {\rho(\x,t) \over  \gamma} )}.
\ee 
Separating  the result into  longitudinal 
terms, i.e.,  parallel to the drift  $\v$, and the transverse terms, 
perpendicular to $\v$, one gets,  
\begin{eqnarray}\label{cole1}
\partial_t \rho(x,\r_\perp,t)  &=&  
D_x  \partial_x^2 \rho(x,\r_\perp,t) +
 D_\perp  \nabla_\perp^2 \rho(x,\r_\perp,t)\\  \nonumber   
&+& {D \over \gamma} [{\bf \nb}_\perp \rho(x,\r_\perp,t) ]^2 
 +{D \over \gamma}  [\partial_x \rho(x,\r_\perp,t) ]^2  \\  \nonumber
  &-& \v \cdot \partial_x \rho(x,\r_\perp,t) +   \gamma (b +   V(x,\r_perp)). 
\end{eqnarray}
Where  for concreteness we take $\v$ along the
$\x$ direction and denote the remaining $d-1$ directions  by $\r_\perp$.

At the 
Burgers's fixed point, i.e., for large $\tilde {h}$,   $x $ scales 
as $r_\perp^{z}$,  and  
$\rho$ scales as $r_\perp^{\alpha}$, where $\alpha = 1/2 $ and $z = 3/2$ 
for the $2+1$ dimensional  realization of the model \cite{FNS}. 
From Eq. (\ref{cole1})
one sees that the terms proportional to   $D_x$ and $(\partial_x \rho)^2  $
scale like $r_\perp^{\alpha - 2 z}$ and $r_\perp^{2 \alpha - 2 z}$ 
respectively, 
which implies that they are irrelevant on large 
length scales. Thus,
the eigenstates of (\ref{cole1}) satisfy the 1+1 Burgers type equation,
where the first order derivative with respect to time is multiplied by
$h$. For more details see \cite{us}

This picture is essentially valid in higher spatial dimensions.  Eq.
(\ref{cole1}) is not specific  to any dimension, so  it gives the 
correct scaling behavior in terms of $\alpha$
and $z$.  The fact that  
$\alpha + z = 2$ in any dimension, where $\alpha$ decreases 
with the dimensionality of the system,  implies that 
the irrelevant terms become even more irrelevant near the Burgers's fixed 
point for more than two spatial dimensions. Thus, one should expect   
that the  delocalized 
ground state in $d+1$ spatial dimensions has the scaling  properties of 
a  directed polymer in $d$ dimensions while extended in the 
direction of the tilt. For $d>3$ this implies that 
for weak disorder the system may be in the weak coupling regime, 
so that the nonlinear terms in (\ref{cole1}) are both irrelevant, and
the system flows into the Gaussian (Edwards-Wilkinson) 
fixed point. As a result,
the wandering exponent of the low-lying states becomes $1/2$, like a classic 
lattice random walk, and  the
energy barriers between these  eigenstates do not scale with the size
of the system. However, above a critical threshold value of the 
disorder strength, the system renormalizes into the strong coupling regime,
and a scenario similar to the $2+1$ dimension case should apply.

\subsection{Energy statistics and finite size effects in the extended phase}

In this 
subsection we  address  the interpolation between the localized and the 
Burgers, directed polymer like,  limit.   
As the  tilt $\tilde{h}$ becomes strong, 
there is a linear response
of the bulk magnetization to the tilt; the average perpendicular 
component of the magnetization ${\bf b_\perp} \equiv {\bf  B_\perp  \phi_0 / 
4 \pi} $ is related to  
the external field   as 
${\bf  b_\perp } =  \sigma({\bf  h_\perp }){\bf  h_\perp }$ where  
$\sigma(\h_\perp) \to 0$ as 
$\h_\perp  \to h_c $, and  $\sigma({\h_\perp}) \to 1$ as $\h_\perp \to \infty$.

For  $\h_\perp > \h_\perp^c$, the vortex line 
proliferates superkinks of the same sign \cite{nel-vin}. 
Thus,  the energy associated with the the existence of the flux line 
at point $\r$ 
is not $U(\r)$, but rather ${\tilde U(\r)} = 
U(\r) \tau(\r,U(\r))$, where $\tau(\r,U(\r))$ is
 the longitudinal length,  
the flux line spent on the point $\r$, which is itself  
a function of the  potential  energy at the point. Since the flux line tends 
to spend more time on strongly attractive points, ${\tilde U(\r)}$ 
is a nonlinear function of $U(\r)$ , 
and the energy distribution function  
$P({\tilde U})$ is no longer 
symmetric; instead, it tends
to emphasize the attractive regions while deemphasizing
the  repulsive ones. Calculation of $P({\tilde U})$ in some limits
 of a  simple model
are presented in Appendix  B.
It is shown that the effective disorder strength diverges as 
$(h - h_c)^{-1/2}$ as $h$ approaches the critical field from above. 

Since actual systems are of finite 
size, it may lead to completely different response to the external drift.
In the directed polymer phase, the disorder strength and the tilt modulus 
lead to  a length scale $L_c$, above which the polymer adapts itself to the 
pinning potential. Below $L_c$, the line remains stiff, since the fluctuations 
of the pinning energy grows only sublinearly with length \cite{rmp}. 
Thus, in the 
Burgers's phase, the low lying  eigenstates may be fluctuating,  
depending  
on the actual length of the sample and the strength of the disorder. 
As $h \to h_c$ from above, the  effective disorder strength diverges, 
such that the ground state is always fluctuating in this limit. 
The various crossover regions for different sized samples are 
illustrated in  Fig. 2.

It is interesting to note that the critical length $L_c$ may be estimated 
by two different ways \cite{nelson}. First,  one may extract the relevant part 
of the   Eq. (\ref{cole1})  (where we omit the  bias $b$, since 
it does not effect the results) 

\begin{minipage}[t]{3.2in}
\epsfxsize=3.2in
\epsfbox{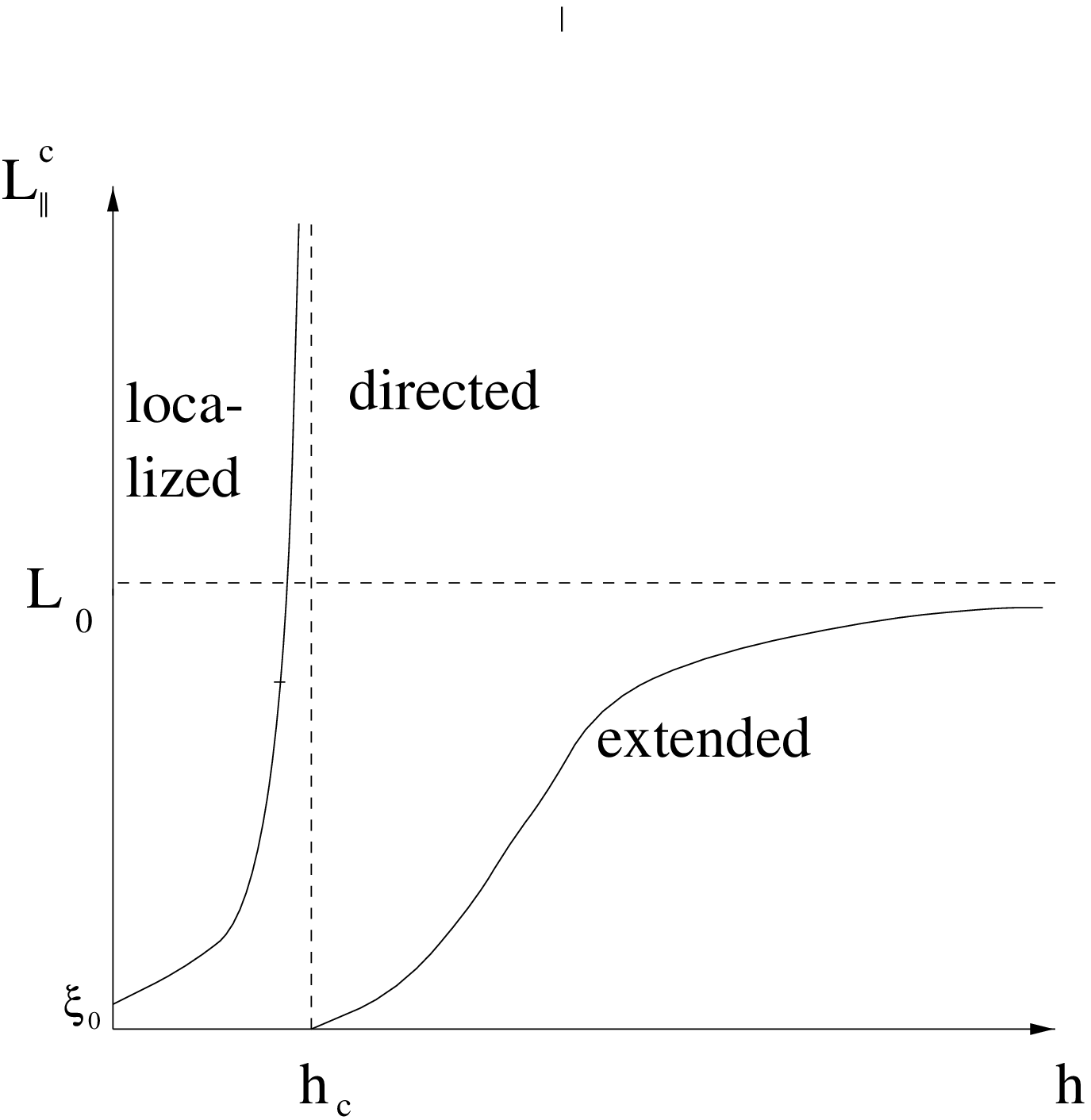}
\begin{small}
FIG.\ 2. 
``Phase diagram'' for the 
different states of the flux line. In order to see the localized phase, the 
longitudinal dimension of the sample should be longer than the localization 
length, which diverges like $(h_c-h)^{-1}$ as $h \to h_c$ from below. 
Above the transition, the potential barriers between two competing 
metastable states scales with the length of the sample only if
$L_\| > Dh/\Delta^2 l_0^3$. As $h \to \infty$, $L_\|^c$  saturates
to its value for point disorder in the $xy$ plane. When the convection
is smaller, the effective disorder becomes stronger; as a result, 
$L_\|^c \sim (h-h_c)^{3/2}$ near the transition.
\end{small}
\vspace{0.2in}
\end{minipage}

\be
h {\partial \u \over \partial x} = 
D_\perp \nabla_\perp^2 \u + {D \over \gamma} (\nb_\perp \u)^2 + \gamma 
V(\r_\perp, \x).
\ee

Division by $h$ and rescaling of the length units by some short length
cutoff $l_0$, i.e., $y = x/l_o$, one gets the dimensionless form,
\be\label{dp}
{\partial \u \over \partial y_\|} = 
{D_\perp \over l_0\ h} \nabla_\perp^2 \u + 
{D \over \gamma  h  l_0 } (\nb_\perp \u)^2 + {\gamma   l_0 \over h} 
 V({\bf y}_\perp, y_\|).
\ee
The Fourier transform of $\u({\bf y}_\perp, y_\|)$,
$\u({\bf k}, \omega)$  satisfy the integral equation,
\begin{eqnarray}
\u&(&{\bf k}, \omega )  = G_0({\bf k}, \omega ) V({\bf k}, \omega ) 
\\ \nonumber 
&+& {D \over \gamma  h  l_0 } \int {d{\bf q} d\Omega \over (2 \pi)^{d}} \ 
[{\bf q} \cdot ({\bf k- q})] \   \u({\bf q}, \Omega ) \   \u({\bf k-q}, \omega - \Omega )
\end{eqnarray}
where 
\be
 G_0({\bf k}, \omega ) = 
{1 \over i \omega -  {D \over \gamma  h  l_0 }{\bf k}^2},
\ee
and $V({\bf k}, \omega)$ is the Fourier transform of the growth rate.

A loop renormalization group  
correction to the dimensionless diffusion constant $
\df =  {D \over \gamma  h l_0 } $ comes from the self energy diagram 
and is of order $g^2 \int dq q^{d-3}$
where the dimensionless coupling constant $g^2 =  {\Delta^2 l_0^3 \over D h}$.
In the  $(2+1)$ dimensional realization of our system, Eq. (\ref{dp}) 
is reduced to  
$1+1$ dimensions, and   this correction  diverges like  $g^2/q$ as the 
incoming momentum vanishes; this implies that 
the transverse  dimension of the system, $L_\perp$, should satisfy
$L_{\perp} >{ D h \over \Delta^2 l_0^3}$ in order to see the effect of the
disorder, i.e., the transition from stiff ground state to a wandering one. 
A trivial
scaling argument then shows that the longitudinal dimension of the
system is given by  $L_{c,\|} = ( D h /  \Delta^2 l_0^3)^{3/2}$.

An alternative way to see the same result is the following:  
consider a $2D$ isotropic homogeneous  tight binding model,
which is a discretization of Eq. (2.3). 
The hopping term, thus, is   $D/l_0^2$, where $l_0$ the lattice constant. 
The energy 
spectrum of this model is simply $E(k_x,k_y) = 
{D \over l_0^2} \  [\cos(k_x l_0 ) + \cos(k_y l_0 )]$, 
where  ${\pi \over L } < k_{x,y} < {\pi}$. Thus, the minimal 
separation between nondegenerate energy levels  is $\Delta E \sim {D  \over
L^2 }$. Introducing the tilt  $\v/D$ pointing in the $\x$ direction, 
leads to a simple shift of $k_x \to k_x + i {h \over D}$, and the 
corresponding states are tilted in the $\x$ direction, without any wandering in
the direction transverse to $\x$. As a result, the eigenstates are now 
complex, and the minimal imaginary distance between energy  levels
is ${D h  \over L_\perp l_0^3}$. 
Now let us introduce disorder into the system. Using first order perturbation
theory one finds that the leading order correction to the energy levels,
is proportional to $\Delta^2$; thus,  we have  the same result,
i.e., that non trivial effects of the disorder may be seen if $L_\perp 
 >  L_{\perp,c} \sim (D h / \Delta^2 l_0^3 )$.

\section*{Acknowledgments}
I would like to thank D. R. Nelson, B. I. Halperin,   K. A. Dahmen and 
 N. Hatano
for most helpful discussions  and comments.This research was supported
by the National Science Foundation through Grant No. DMR94--17047, 
and by the Harvard Materials Research Science and Engineering 
Laboratory through Grant No. DMR94--00396,  
Bar-Ilan University and a Rothschild  Fellowship.

\appendix 

\section {Tail states statistics}

In this appendix we  calculate the density of states (DOS) at the 
tails of the band, i.e., near the ground state. We then use the
results  to study the decay of these states to the ground state. The model
we use is a discretization of Eq. (2.3), where we neglect 
the constant bias of the growth rate $b$, since it does not effect the 
statistics of the DOS. The discussion below assumes localization of the 
wavefunction of the  relevant part of the eigenvalue spectrum, hence, 
the tilt term   $\v$  does not change the spectrum; it effects
the decay rate via the modified  localization length, as discussed
in Section 3b.

We consider, thus,  a model 
 of a $d$ dimensional configuration  space tilted by square blocks of 
length $l_0$, where the potential energy 
in each of these blocks is a  constant number taken from a  
square distribution 
between $0$ and $\Delta$ (the choice of these limits is only for 
the sake of convenience. 
since the  results are independent of the potential  bias $b$, on the other
hand, other probability measures may give different results. In particular, 
for any unbounded distribution,  such as Gaussian 
distributions, the spectrum of $\H$ is unbounded from above, and there 
is no delocalization transition in the sense we discuss here). 
The dimensionless Hamiltonian  is :
\be\label{hem}
\H =  -  D \nabla^2 + V({\bf x}),.
\ee
It is easy to see that the DOS function $\rho(\E)$  in this model is  bounded
from below  by  zero, and $\rho(\E) \to 0$ as $\E \to 0$. The tail of the 
density of states is determined by the range of parameters in which
rare events, such as large spatial regions with low potential energy, 
determine the DOS.  

Let us estimate these fluctuations in the following way:  the probability to 
find a hypersphere of radius $R$ which  contains 
only blocks of potential energy 
less then $V_0$ is   
\be
P(R, V < V_0) \sim \exp([R/l_0]^d \ln [V_0/\Delta]).
\ee
The ground state energy of these fluctuations is given approximately by
\be 
\E_0 \sim  V_0 + {D \over R^2 l_0^2}
\ee

so that the probability to get the energy between  $\E$ and $\E + d\E$ 
using a sphere of radius $R$
is,
\be
p(R, \E) \sim \exp( \left({R \over l_0} \right)^d \ln (\E - 
{D \over R^2} ))
\ee
This expression is well defined for $ {\sqrt {D / \E}} \leq R < \infty  $.
Optimizing $p(R, \E)$ with respect to $R$ gives, up to logarithmic 
corrections, a  maximum at the lower limit  
$R^* =  \sqrt{ D/\E}$, so that as $\E \to 0$
from below, 
\be\label{qqq}
P(\E) \sim \exp(-(\df/(-\E))^{d/2}).
\ee

We assume that the DOS is proportional to $P(\E)$, i.e., at the tail
of the distribution we have  
$g(\E) \sim g_0 P(\E)$, where $g_0$ is some normalization
proportional to the DOS at the middle of the band. 

Let us consider now the  tight binding analog of the above model. 
The on site potential is now $w_i$, taken from a  square distribution between
zero and $\Delta$. The  Hamiltonian is
\be
H = {t \over 2} \sum_\x \sum_{\nu = 1}^d  
(a^\dagger_{\x + \bf{e}_\nu} a_\x + h.c.) - \sum_i v_i a^\dagger_\x a_\x
\ee
where $a^\dagger_\x, a_\x$ are boson creation and annihilation operators,
$\bf{e}_\nu$ are unit lattice vectors, and the hopping element $t$ is 
normalized by $D$.   

The eigenenergies of this Hamiltonian are  bounded, 
$-t < \epsilon_n < \Delta+t$.
The states at the tails correspond  to a rare spatial fluctuations 
of $v_i$'s; in general, the chance to find such fluctuations (e.g., a region
of radius $R$ in which the on site potential is less then $V$)  is 
the same as in the previous model. The  energy spectrum of such fluctuation
is given approximately by
\be
\epsilon =  V + t \sum_{\nu = 1}^d \cos(k_\nu)
\ee
where $k_{min} \sim 1/R$, thus, the tail states obey the relation
\be
\epsilon \sim  V -  t + t/R^2
\ee
so that the  result (\ref{qqq}) is  applicable 
here also, with the energy  measured from the lower bound of the band.

\section{Magnetic response  at large tilt   }

In this Appendix we study the response of the bulk   
magnetization $B_\perp$ 
to the tilt of the external field. In the 
localized phase there is no linear response, i.e., no bulk tilt of the
vortex as a result of the external perpendicular magnetic field. 
Thus, our discussion here is relevant only in the extended phase. 

It is important to note that for an {\it unbounded} distribution 
of defects strength, such as a  Gaussian distribution, one flux line is 
always localized. For a given tilt, there is always  
a strong enough defect such
that the localization length is small enough and the  tilted state 
is localized. A delocalization of one flux line may occur only if the 
distribution of the defects is bounded. We will take the quenched 
random potential from a  distribution of width $\Delta$, so that the 
minimal localization length is of order $1/\sqrt{\Delta}$. The tilt
is assumed to be strong such that all the states are localized.

 The typical situation, though, is that the flux line ``lives'', for a while, 
on defects which are quite strong with respect to their neighbors, i.e., 
there are no stronger defects in the vicinity of the strong one.
After a while, 
however, the flux line is going to jump to another defect which is 
weaker, in order to satisfy the tilt term in the free energy. This process 
may be described by the optimization of the free energy of the ``local jump''
\be\label{fe}
\delta F = -{\Delta \tau \over (1+ \Delta g(\mu) R^d)} 
- \int {\bf H \cdot B}(z) dz, 
\ee
where $\tau$ is the length that the flux line spend 
on the defect.
The first term in (\ref{fe}) indicates  that there is a decrease of the
free energy due to the fact that the flux line lives on the defect. This 
decrease is proportional to the potential energy difference between the 
current site and the site the flux line jumps to; it is bounded from above
by $\Delta$ while  it  decreases to zero when  the jump distance, $R$, 
is much bigger than the density of states per unit area, since then the 
flux line may find another strong defect. The denominator of the 
first term in (\ref{fe}) 
reflects the simplest interpolation between these two limits. 
The term  $ {\bf H \cdot B}$
indicates the free energy per unit length decrease since   the 
flux line average  direction ${\bf B}$ tends to
be in the direction of the external magnetic field ${\bf H}$. In terms of the 
``local jump'', this term takes the form
\be
{\bf H \cdot B} = {H \phi_0 \over 4 \pi}(\tau^2 + R^2)^{1/2}
\cos\{atg[{H_\perp \over H_|}] - atg[{R \over \tau}]\}.
\ee   

Now let us calculate the effective tilt of the flux line, i.e., the response 
of the internal magnetic field ${\bf B}$  to the external field ${\bf H}$. 
For ${\bf H}$ in the direction of the defects, ${\bf B}$ is parallel to ${\bf  
H}$; this is also
true if  ${\bf H}$ is perpendicular to the defects. Small tilt of 
 ${\bf H}$ from the defects direction is not enough to tilt the 
flux line, so that there is no response of the internal magnetic field (except
near  the surface, see \cite{hat}). Above the critical tilt there is 
linear response of the flux line, i.e., $B_\perp = \sigma(H_\perp) H_\perp$, 
where $\sigma(H_\perp) \to 0$ as 
 $H_\perp \to H^c_\perp$ and $\sigma(H_\perp)$ approaches 
$1$ as $H_\perp \to \infty$.
Using our local jump model, near $H_\perp^c$ i.e, where  $H_\perp << H_\|$ and
$R << \tau$; near the critical tilt $R \to 0$, so that  we may  
assume that $R^d << \Delta g$. $\delta F$  then  takes the form
\be
\delta F = - \Delta \tau + {H \phi_0 \over 8 \pi} \tau [{H_\perp \over H_\|}
- {R \over \tau}]^2.
\ee
Optimizing with respect to $\tau$ one gets 
\be
{H \u_0 \over 8 \pi} [({H_\perp \over H_|})^2
- ({R \over \tau})^2]  \approx \Delta.
\ee
Since, in this limit $B_\perp  \sim  
R / \tau $ and $H_\perp \sim {H_\perp / H_\|}$, 
and using the definition
$\ept ={H \phi_0 \over 4 \pi}$, we have 
\be
B_\perp^2 = H_\perp^2 - {\Delta \over \ept }
\ee
i.e., below  $ H_\perp^c = \sqrt{\Delta/{\tilde\epsilon}}$ there is no 
response to the external tilt while above $ H_\perp^c$ 
$ B_\perp \sim ( H_\perp -    H_\perp^c)^\beta $ where $\beta = 1/2$. 
This analysis is valid for the region in which both $B_\perp$ and 
$H_\perp$ are much smaller than one, and the resulting $R$ is small
compared to  $(g \Delta)^{1/d}$.

In the opposite limit, where the external magnetic field is 
perpendicular to the defects, $R$ is much bigger than both  $\tau$
and  $(g \Delta)^{1/d}$, thus the first term is negligible  and the 
flux line is parallel to the external field. 

The above treatment was for one jump, i.e., it neglects global 
effects of the defects. For the actual $B_\perp$, one should take the 
typical defect instead of the extreme one. However, we may use the same 
analysis in order to get the ``time'', $\tau$, spent 
by the flux line on a  particular 
defect, by  dividing the radius of the defect 
by  the ``local velocity'', i.e.,
the associated  
perpendicular part of the internal  magnetic field, associated 
with it. The result is
\be
\tau (U) \sim {R_0 \over  B_\perp} \sim {R_0 \over  \sqrt{ H_\perp^2 - 
 [{U \over {\tilde\epsilon}}]}}
\ee
where $R_0$ is the radius of the defect. Thus, the effective energy 
associated with a defect of potential energy $U$ is tilt dependent;
as one approaches the critical tilt for this defect, $H_\perp \to H_\perp^c + 
\delta$,  the effective energy $E_{eff}
= U \tau(U)$ diverges as $1/\sqrt{\delta}$.   
(This is in agreement
with the result for $h_c$ from the localized limit, i.e., by equating the 
typical inverse localization length $\sqrt{\Delta/D} $ to the $h/D$ term in the
gauge transformation \cite{me}).

\end{multicols}{2} 
\end{document}